\documentclass[%
reprint,
superscriptaddress,
showpacs,
showpacs,preprintnumbers,
amsmath,amssymb,
aps,
]{revtex4-1}

\usepackage[dvipdfmx]{graphicx}
\usepackage{dcolumn}
\usepackage{bm}
\usepackage{multirow}
\usepackage[mathlines]{lineno}

\usepackage{color}

\begin{document}

\preprint{APS/123-QED}

\title{Anomalous Hall effect triggered by pressure-induced magnetic phase transition in $\alpha$-Mn}

\author{Kazuto~Akiba}
\email{akb@okayama-u.ac.jp}
\affiliation{
Graduate School of Natural Science and Technology, 
Okayama University, Okayama 700-8530, Japan
}

\author{Kaisei~Iwamoto}
\affiliation{
Graduate School of Natural Science and Technology, 
Okayama University, Okayama 700-8530, Japan
}

\author{Takaaki~Sato}
\affiliation{
Graduate School of Natural Science and Technology, 
Okayama University, Okayama 700-8530, Japan
}

\author{Shingo~Araki}
\affiliation{
Graduate School of Natural Science and Technology, 
Okayama University, Okayama 700-8530, Japan
}

\author{Tatsuo~C.~Kobayashi}
\affiliation{
Graduate School of Natural Science and Technology, 
Okayama University, Okayama 700-8530, Japan
}

\date{\today}

\begin{abstract}
Recent interest in topological nature in condensed matter physics has revealed the essential role of Berry curvature in anomalous Hall effect (AHE). 
However, since large Hall response originating from Berry curvature has been reported in quite limited materials,
the detailed mechanism remains unclear at present.
Here, we report the discovery of a large AHE triggered by a pressure-induced magnetic phase transition in elemental $\alpha$-Mn.
The AHE is absent in the non-collinear antiferromagnetic phase at ambient pressure, 
whereas a large AHE is observed in the weak ferromagnetic phase under high pressure despite the small magnetization of $\sim 0.02 \mu_B$/Mn.
Our results indicate that the emergence of the AHE in $\alpha$-Mn is governed by the symmetry of the underlying magnetic structure,
providing a direct evidence of a switch between a zero and non-zero contribution of the Berry curvature across the phase boundary.
$\alpha$-Mn can be an elemental and tunable platform to reveal the role of Berry curvature in AHE.
\end{abstract}

\maketitle

\section{Introduction}
Anomalous Hall effect (AHE) in systems with broken time-reversal symmetry
is one of the fundamental transport phenomena in condensed matter physics \cite{Hall_1880}.
In general, the Hall resistivity $\rho_{yx}$ is represented as $\rho_{yx}=\rho_{yx}^N+\rho_{yx}^A$ \cite{Smith_1929, Pugh_1950}.
Here, $\rho_{yx}^N$ is the normal component due to the Lorentz force,
whereas $\rho_{yx}^A$ represents the anomalous component observed in an magnetically ordered phase,
which becomes empirically larger when the system has a larger spontaneous magnetization ($M$).
Conventionally,
it is widely acknowledged that spin-dependent scattering processes in the presence of
$M$ (so called ``extrinsic'' origins) result in the AHE \cite{Smit_1955, Smit_1958, Berger_1970}.
On the other hand, a recent interest in topological nature in condensed matter physics has provided insight on the ``intrinsic'' origin of the AHE \cite{Karplus_1954},
which is re-interpreted to be Hall response due to the Berry curvature in the momentum space \cite{Ohgushi_2000, Onoda_2002, Jungwirth_2002, Fang_2003, Haldane_2004}.
The anomalous Hall conductivity $\sigma_{xy}^A$ is represented by the Kubo formula as \cite{Jungwirth_2002, Fang_2003}
\begin{equation}
\sigma_{xy}^{A}=-\frac{e^2}{\hbar}\sum_{n}\int \frac{d\bm{k}}{(2\pi)^3}f[\epsilon_n(\bm{k})] b_{n}^{z}(\bm{k}),
\label{eq_AHC}
\end{equation}
where $e$, $\hbar$, $n$, $\bm{k}$, $\epsilon_n(\bm{k})$, and $f$ represent the elemental charge, reduced Planck constant, band index, wavevector, eigenvalues of the Hamiltonian, and Fermi–-Dirac distribution function, respectively.
$b_{n}^{z}(\bm{k})$ represents the $z$-component of the Berry curvature \cite{Berry_1984},
which acts like a magnetic field in the momentum space.
The norm and direction of $\bm{b}_{n}(\bm{k})$ are determined only by the Bloch state of the corresponding energy band.
As Eq. (\ref{eq_AHC}) presents, $\sigma_{xy}^{A}$ becomes non-zero
when the integration of the Berry curvature over the occupied states in the momentum space remains finite, regardless of the net $M$ or scattering events.

This mechanism is expected to cause a large Hall response in antiferromagnetic (AFM) systems with certain symmetry conditions.
A cubic non-collinear antiferromagnet Mn$_3$Ir, whose Mn sublattice can be regarded to as stacked kagome lattice along the [111] direction, is theoretically expected to show large anomalous Hall conductivity $\sigma_{xy}^{A}\sim200  \Omega^{-1}$ cm$^{-1}$ for its triangular spin order \cite{Chen_2014}.
This value is not at all inferior to that in elemental ferromagnet Fe ($1000 \Omega^{-1}$ cm$^{-1}$) \cite{Yao_2004} despite the absence of net magnetization.
Although this prediction has not fully been verified, a recent experiment on Mn$_3$Ir thin film reported anomalous Hall conductivity as large as $\sigma_{xy}^{A}\sim40  \Omega^{-1}$ cm$^{-1}$ at room temperature \cite{Iwaki_2020}.
Similar large intrinsic AHE has been theoretically expected  in hexagonal non-collinear antiferromagnets Mn$_3$Sn and
Mn$_3$Ge \cite{Kubler_2014}, which have an inverse triangular spin structure with quite small ferromagnetic component.
Actually, subsequent experiments \cite{Nakatsuji_2015, Nayak_2016, Kiyohara_2016} revealed that
the anomalous Hall conductivity is strongly anisotropic, and reaches approximately $\sigma_{xy}^{A}\sim150  \Omega^{-1}$ cm$^{-1}$ and $400  \Omega^{-1}$ cm$^{-1}$ in Mn$_3$Sn and
Mn$_3$Ge, respectively.
Interestingly, recent progress has revealed possible large AHE even in collinear antiferromagnets \cite{Smejkal_2020, feng2020observation, Naka_2020}.
In this context, the search for a large intrinsic Hall response has attracted attention not only to understand the long-standing issue in condensed matter physics
but also to identify an application for a novel sensor and memory device.
However, such a large intrinsic AHE is reported in quite limited materials at present.
Thus, a model material that enables to flexibly control the electronic structure by external parameters is desired.

\begin{figure}[]
\centering
\includegraphics[]{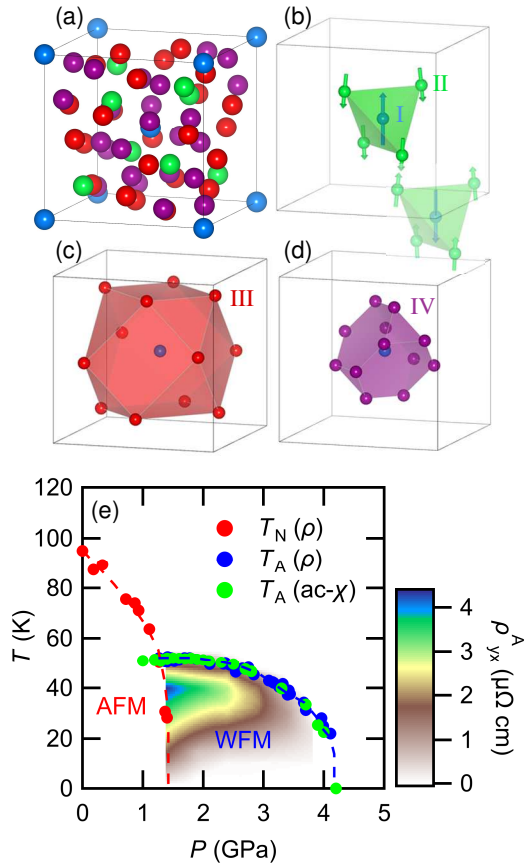}
\caption{
(a) Crystal structure of $\alpha$-Mn. Blue, green, red, and purple spheres represent sites I, II, III, and IV, respectively.
(b, c, d) Atomic configurations around site I.
For sites I and II, the orientation of the magnetic moments in the antiferromagnetic phase
is also illustrated based on Ref. \cite{Lawson_1994}.
(e) $P$-$T$ phase diagram of $\alpha$-Mn reprinted from Ref. \cite{Sato_2020}. $T_N$ and $T_A$ represent the antiferromagnetic transition temperature and pressure-induced phase transition temperature, respectively.
The red and blue symbols are from resistivity measurements and the green symbols are from ac-susceptibility measurements.
The color plot displays the anomalous Hall resistivity $\rho^A_{yx}$ obtained by the present study (see main text).
\label{fig_1}}
\end{figure}

Here, we report the discovery of a large AHE in $\alpha$-Mn,
a stable form of elemental Mn at room temperature and ambient pressure.
$\alpha$-Mn forms a body centered cubic (bcc) structure that consists of 58 atoms in the bcc unit cell with 4 non-equivalent Mn sites
referred to as sites I, II, III, and IV [Figs. \ref{fig_1}(a)--(d)].
It belongs to the non-centrosymmetric space group $I\bar{4}3m$.
$\alpha$-Mn is known to exhibit an AFM transition at $T_{\rm N}=95$ K \cite{Miyake_2007, Takeda_2008},
in which the magnetic moments on each Mn site
(1.9, 1.7, 0.6, and 0.2 $\mu_{\rm B}$ for sites I, II, III, and IV, respectively)
form a non-collinear AFM spin structure \cite{Yamada_1970, Yamagata_1972, Lawson_1994, Hobbs_2003, Pulkkinen_2020}.
Here, $\mu_{\rm B}$ represents a Bohr magneton.
$T_N$ is rapidly suppressed by the application of pressure, and another pressure-induced phase characterized by the transition temperature $T_A$ appears above 1.4 GPa,
which results in a double-stage structure in the pressure--temperature ($P$-$T$) phase diagram as shown in Fig. \ref{fig_1}(e).
Recently, a significant increase in the ac-susceptibility in this high-pressure phase was reported \cite{Sato_2020}; 
however, the details of the magnetic structure remains uncertain.
In the present study, we identified that the high-pressure phase has a weak ferromagnetic (WFM) nature with quite a small magnetization.
An significant jump of $\rho_{yx}$, which is ascribed to be the AHE, was observed only within the WFM phase.
Our results indicate that the occurrence of the AHE is determined by the symmetry of the underlying magnetic structure,
which is a remarkable evidence of the switch between the zero and non-zero contributions of the Berry curvature
across the phase boundary.

\section{Experimental methods}
Single crystals of $\alpha$-Mn were synthesized by the Pb-flux method.
Mn (99.999\%) and Pb (99.9999\%) with a molar ratio of 2:98 were placed in an alumina crucible and sealed in a quartz ampoule with argon gas.
After the mixture were initially heated to $800\ {}^\circ\mathrm{C}$, the melt was cooled to $320\ {}^\circ\mathrm{C}$ for over 300 h.
Then, the flux was removed using a centrifuge separator.
A picture of the as-grown crystal is shown in the inset of Fig. \ref{fig_2}.
The top surface of this sample was confirmed to be (101) plane by means of X-ray diffraction analysis.
It was shaped by mechanical polishing into rectangular cube ($\sim0.5 \times 0.1 \times 0.1$ mm$^3$) for precise determination of resistivity.
The resistivity of the sample at ambient pressure was
197 $\mu \Omega$ cm and 11.6 $\mu \Omega$ cm at 300 K and 1.7 K, respectively,
and the resulting residual resistivity ratio was 17.

The electrical transport measurements under high pressure were performed
by indenter-type pressure cell ($P < 4$ GPa) \cite{Kobayashi_2007}.
Temperature dependence of the resistivity at zero-field was measured by using a gas-flow-type optical cryostat (Oxford Instruments, $T > 2$ K)
and by a standard four-terminal method with 2400 sourcemeter and 2182A nanovoltmeter (Keithley Instruments).
The effect of thermal electromotive force by temperature gradient was removed by inversion of the current ($I$) direction.
Magnetoresistivity $\rho_{xx}$ and Hall resistivity $\rho_{yx}$ in a static magnetic field were measured using a superconducting magnet (Oxford Instruments, $B < 8$ T)
and variable temperature insert (Oxford Instruments, $T > 1.6$ K).
$\rho_{xx}$ and $\rho_{yx}$ were measured on an identical sample shown in the inset of Fig. \ref{fig_2}
by a standard four-terminal method with LR-700 AC resistance bridge (Linear Research).
In $\rho_{xx}$ and $\rho_{yx}$ measurements, $B$ were applied parallel to [101] direction,
and the current were injected within the (101) plane.
All data in Fig. \ref{fig_3}(c) and \ref{fig_3}(d) are anti-symmetrized as a function of $B$ to remove the effect of misalignment of the voltage contacts,
whereas data shown in Fig. \ref{fig_4}(a) is raw data without anti-symmetrization to show the finite hysteresis.
Silver epoxy and paste (Epo-tek H20E and Dupont 4922N) were used to form electrical contacts.

Magnetization measurements under high pressure were performed by ceramic-anvil pressure cell ($P < 3$ GPa)
\cite{Tateiwa_2011}.
Magnetization was measured by a SQUID magnetometer (MPMS, Quantum Design, $B < 1$ T).
A single crystal was mechanically shaped into rectangular cube and placed in the hole of NiCrAl gasket together with a small piece of Pb pressure marker (the sample space is 0.5 mm diameter and 0.5 mm height).
Since the signal from the sample was relatively smaller than that from the pressure cell itself,
we measured the background magnetization from the pressure cell without a sample,
and subtracted it from the net signal.
The volume of the rectangular sample was estimated by measuring the length of the sides
(Data in Fig. \ref{fig_3}(a) and \ref{fig_3}(b) were measured on separate samples with their volumes of
$8.3\times10^{-6}$ cm$^{3}$ and $1.3\times10^{-5}$ cm$^{3}$, respectively).
The pressure in the sample space was determined by the superconducting transition temperature of Pb at zero-field.

In all high-pressure measurements mentioned above, Daphne oil 7474 \cite{Murata_2008} was used as a pressure
medium.

\section{Results and discussion}

\begin{figure}[]
\centering
\includegraphics[]{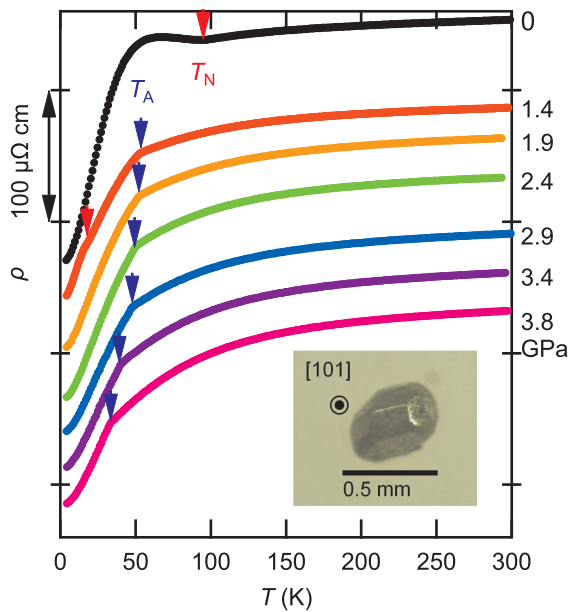}
\caption{
Temperature dependence of the resistivity at zero-field $\rho$ at several pressures.
Each curves are vertically shifted for clarity.
Antiferromagnetic transition temperature $T_N$ and pressure-induced phase transition temperature $T_A$ are indicated by red and blue arrow, respectively.
The inset shows a single crystal of $\alpha$-Mn utilized in the present study.
\label{fig_2}}
\end{figure}

In Fig. \ref{fig_2}, we firstly show the temperature dependence of the resistivity at zero-field ($\rho$) at several pressures.
$I$ was injected in the (101) plane. 
We observed clear anomaly at the transition temperature of the antiferromagnetic phase ($T_N$) and pressure-induced phase ($T_A$), which is denoted by red and blue arrow, respectively.
We confirmed that the pressure dependences of $T_N$ and $T_A$ agree with the previous result \cite{Sato_2020}.

\begin{figure}[]
\centering
\includegraphics[]{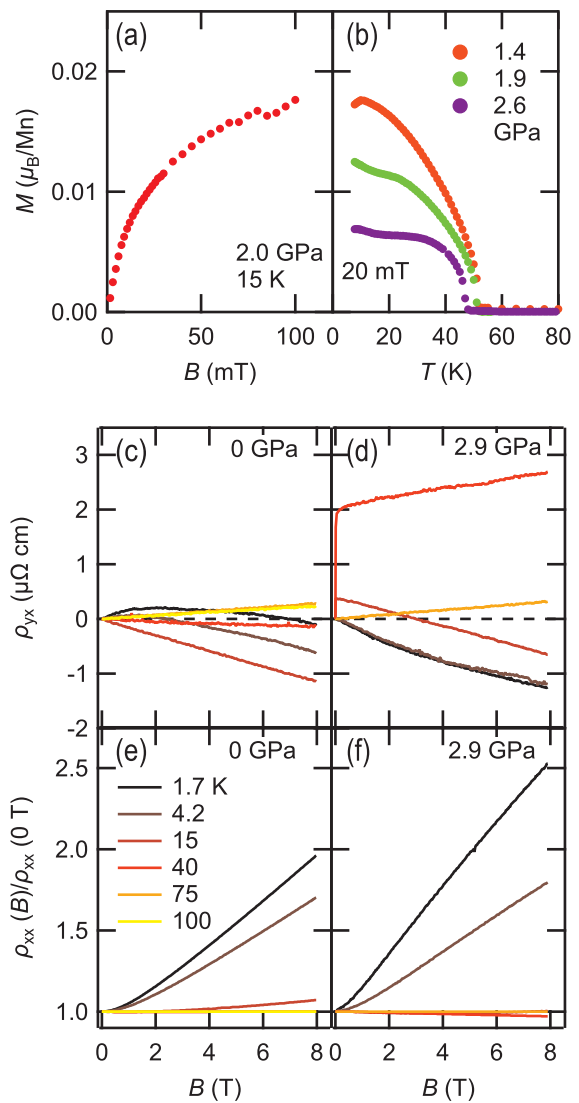}
\caption{
(a) Magnetization $M$ at $P=2.0$ GPa and $T= 15$ K.
(b) Temperature dependence of $M$ at 1.4, 1.9, and 2.6 GPa. External magnetic field of 20 mT was applied during the measurements.
Hall resistivity $\rho_{yx}$ at ambient pressure (c) and at 2.9 GPa (d) with $B \parallel [101]$ and $I \perp B$.
Magnetoresistivity normalized by zero-field value [$\rho_{xx}(B)/\rho_{xx}(\mathrm{0 T})$]
at ambient pressure (e) and at 2.9 GPa (f) with $B \parallel [101]$ and $I \perp B$.
\label{fig_3}}
\end{figure}

Next, we present the magnetic properties in the pressure-induced phase.
As shown in Fig. \ref{fig_3}(a), a small magnetization $M\sim 0.02 \mu_B$/Mn at $B=100$ mT and $P= 2.0$ GPa was identified for the first time,
indicating the WFM nature of this phase.
As shown in Fig. \ref{fig_3}(b), $M$ in the WFM phase is suppressed by further application of pressure.

Subsequently, the Hall resistivity ($\rho_{yx}$) in magnetic fields along [101] direction is focused.
Figure \ref{fig_3}(c) shows $\rho_{yx}$ at ambient pressure.
The non-linear $B$-dependence and sign inversion are assumed to be a trivial contribution in a system in which electrons and holes with different mobilities coexist.
$\rho_{yx}$ exhibits a remarkable non-linearity at low temperatures, whereas it becomes almost linear above 15 K.
At ambient pressure, $\rho_{yx}$ does not display any qualitative difference when $T$ passes through $T_N=95$ K.
On the other hand, a remarkable jump in $\rho_{yx}$ was observed in the WFM phase, as shown in Fig. \ref{fig_3}(d).
This strongly indicates that $\rho_{yx}$ acquired an anomalous Hall resistivity $\rho_{yx}^A$
associated with the pressure-induced magnetic phase transition.
The weak $B$-dependence of $\rho_{yx}$ after the jump is considered to be due to the normal components $\rho_{yx}^N$,
as the qualitative trend is identical to that described in Fig. \ref{fig_3}(c).
We simultaneously performed magnetoresistivity measurements,
whose results are shown in Fig. \ref{fig_3}(e) and (f).
Figure \ref{fig_3}(e) shows magnetoresistivity normalized by zero-field value [$\rho_{xx}(B)/\rho_{xx}(\mathrm{0 T})$] 
at ambient pressure. 
Positive non-saturating magnetoresistance effect of $\rho_{xx}(B)/\rho_{xx}(\mathrm{0 T})\sim 2$ was observed at 1.7 K, and it rapidly suppressed as temperature increases. 
As with Hall resistivity, magnetoresistivity also did not show any qualitative difference when the temperature got across $T_N=95$ K. 
Figure \ref{fig_3}(f) shows $\rho_{xx}(B)/\rho_{xx}(\mathrm{0 T})$ at 2.9 GPa. 
Compared with Fig. \ref{fig_3}(e), the positive non-saturating magnetoresistance at 1.7 K is slightly enhanced to 2.5. 
At intermediate temperature, small negative magnetoresistance was observed ($\sim$3\% decrease at $B=8$ T and $T=40$ K),
whose origin is unclear at present.

\begin{figure}[]
\centering
\includegraphics[]{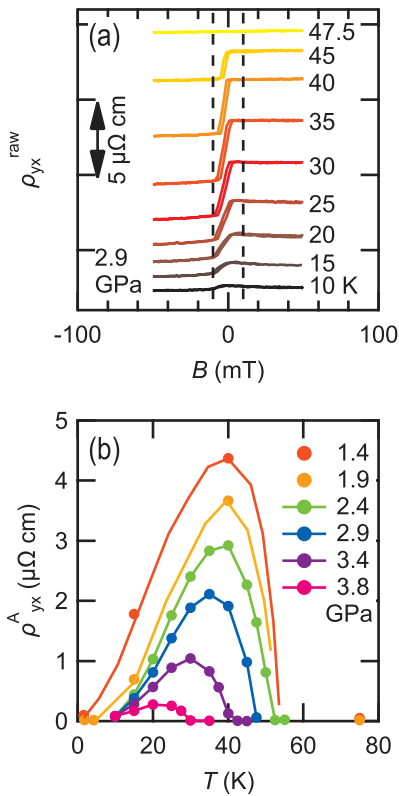}
\caption{
(a) Raw signal of Hall resistivity $\rho_{yx}^{\mathrm{raw}}$ at 2.9 GPa at several temperatures.
$\rho_{yx}^{\mathrm{raw}}$ is not anti-symmetrized as a function of $B$ and vertically shifted for clarity.
Vertical broken lines indicate $B=\pm 10$ mT.
(b) Temperature dependence of the anomalous Hall resistivity ($\rho^A_{yx}$) at several pressures.
\label{fig_4}}
\end{figure}

We further focus on the anomalous Hall part $\rho_{yx}^A$ and the detailed temperature dependence.
Figure \ref{fig_4}(a) shows $\rho_{yx}^{\mathrm{raw}}$ within $\pm 50$ mT at 2.9 GPa.
Note that $\rho_{yx}^{\mathrm{raw}}$ shown in Fig. \ref{fig_4}(a) is not anti-symmetrized as a function of $B$.
A jump in the vicinity of the zero-field occurs as the temperature increases and approaches the maximum at 35 K.
Subsequently, it is suppressed as the temperature increases and vanishes with $T_A \sim 47$ K as the boundary.
The sign inversion with a finite hysteresis loop can be realized by the application of $|B|<10$ mT,
indicating quite a small switching field.
This switching field is smaller than that reported for Mn$_3$Sn and Mn$_3$Ge, typically 10--100 mT \cite{Nakatsuji_2015, Kiyohara_2016}.
Within $\pm 50$ mT, 
the contribution of $\rho_{yx}^N$ is negligibly smaller than the magnitude of the jump.
Therefore, $\rho_{yx}^A$ can be defined as [$\rho_{yx}^{\mathrm{raw}}$ (50 mT)$-\rho_{yx}^{\mathrm{raw}}$ ($-$50 mT)]/2.
The temperature dependence of $\rho_{yx}^A$ at several pressures were determined in the same manner.
As summarized in Fig. \ref{fig_4}(b) and color plot in Fig. \ref{fig_1}(e),
$\rho_{yx}^A$ in the WFM phase can be extensively controlled by $P$ and $T$.
$\rho_{yx}^A$ reaches the maximum near the boundary between the AFM and the WFM phases,
and subsequently decreases as the pressure increases.

Here, a possible magnetic structure realized in the WFM phase is discussed.
In the AFM phase, the moments at sites I and II are relatively larger than those at the other sites,
and thus, these two major sites are focused for simplicity.
As shown in Fig. \ref{fig_1}(b), site I, whose moment is parallel to [001] direction, is included in a tetrahedron formed by site II.
The moments of site II are directed nearly opposite but slightly deviate from [001].
The moments owing to the tetrahedron located at the corner and center of the bcc unit cell cancel each other, which results in a non-collinear AFM phase at the ambient pressure.
As the magnetization in WFM phase is small compared to those of each Mn sites in AFM phase,
a simple ferromagnetic order, in which all of the moments at each site align the same direction should be excluded from possible candidates.
Since previous theoretical calculation \cite{Hobbs_2003} suggests compression-dependent change of the AFM configuration,
it is reasonable to regard the WFM phase as a magnetic order with a slight change from the AFM phase.
One of possible candidates for WFM phase can be considered, in which the tetrahedra at the center and corner of the bcc unit cell ferromagnetically align.
Considering the observed small magnetization, the moments by sites I and II should mostly cancel each other, whereas the residual magnetization can emerge as the cancellation between the center and the corner of the bcc lattice is disabled in this configuration.
The above picture is proposed as a possible candidate of the WFM phase,
which should be clarified by further studies in the future.

Subsequently, the AHE observed in the WFM phase is discussed.
As mentioned above, the AHE can be caused by both extrinsic and intrinsic origins.
In the present case,
it is unlikely to occur such a drastic enhancement of impurity scattering effect by pressure only in the WFM phase.
Thus, our results strongly indicate that
the WFM phase possesses a large
non-trivial contribution of the Berry curvature that do not cancel out by the integration in Eq. (\ref{eq_AHC}),
in contrast with the AFM phase.
In the following, we quantitatively demonstrate that the large AHE observed in the WFM phase originates from the intrinsic effect.

\begin{figure}[]
\centering
\includegraphics[]{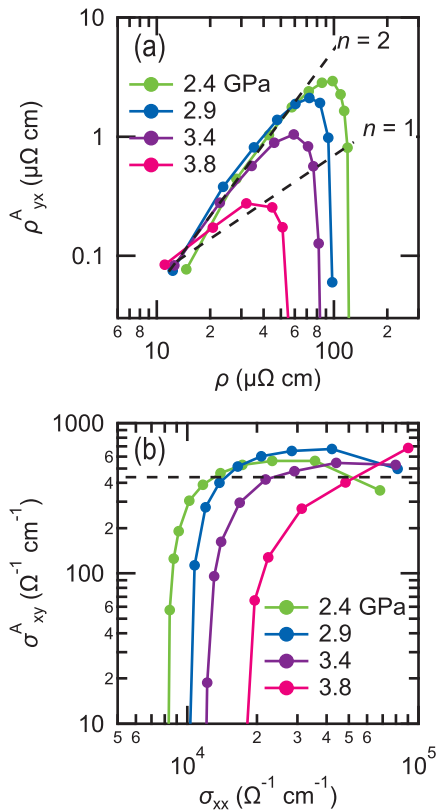}
\caption{
(a) Log-log plot of the anomalous Hall resistivity $\rho^A_{yx}$ as a function of resistivity $\rho$.
Broken lines represent the slope of $n=2$ and $n=1$ cases assuming $\rho^A_{yx} \propto \rho^n$.
(b) Anomalous Hall conductivity ($\sigma^A_{xy}$) as a function of conductivity ($\sigma_{xx}$).
Horizontal broken line indicates $\sigma_{xy}^A=e^2/(ha)=440$ $\Omega^{-1}$ cm$^{-1}$ using $a=8.841$ $\mathrm{\AA}$ (see main text).
\label{fig_5}}
\end{figure}

According to a previous study \cite{Karplus_1954},
the intrinsic mechanism predicts that $\rho_{yx}^A\propto \rho^2$, where $\rho$ represents the resistivity at zero-field.
Figure \ref{fig_5}(a) shows the variation of $\rho_{yx}^A$ as a function of $\rho$.
The traces at 2.4, 2.9, and 3.4 GPa are better applied to the quadratic relation rather than linear one,
which is consistent with the intrinsic mechanism.
At 3.8 GPa in Fig. \ref{fig_5}(a), $\rho_{yx}^A$ deviates from the quadratic relation
and approaches $\rho$-linear relation.

To obtain further insight, we discuss the anomalous Hall conductivity $\sigma_{xy}^A$,
which is connected with the Berry curvature by Eq. (\ref{eq_AHC}).
As described above, the intrinsic $\sigma_{xy}^A$ should be independent of $\tau$ as it depends only on the Bloch state.
Figure \ref{fig_5}(b) shows $\sigma_{xy}^A=\rho_{yx}^A/(\rho^2+{\rho^{A}_{yx}}^2)$ as a function of $\sigma_{xx}=\rho/(\rho^2+{\rho^{A}_{yx}}^2)\sim 1/\rho \propto \tau$.
Although $\sigma_{xx}$ varies nearly an order of magnitude, $\sigma_{xy}^A$ remains almost constant except at 3.8 GPa,
indicating that the AHE is irrelevant to $\tau$.
In the present understanding of AHE, the dominant mechanism varies depending on the relationship between Fermi energy $E_F$, spin-orbit interaction energy $\epsilon_{SO}$, and relaxation time $\tau$ of the system \cite{Onoda_2006, Miyasato_2007, Nagaosa_2010}.
The skew scattering \cite{Smit_1955, Smit_1958} can be dominant in a super clean case ($\hbar/\tau \ll \epsilon_{SO}$), and decays as $\hbar/\tau$ increases compared to $\epsilon_{SO}$.
In the intermediate scattering strength ($\epsilon_{SO} < \hbar/\tau < E_F$),
the $\sigma_{xy}^A$ is mainly governed by the Berry curvature, and takes almost universal value $e^2/(ha)\sim$ 100-1000 $\Omega^{-1}$ cm$^{-1}$,
where $a$ is a lattice constant.
This value is qualitatively explained by assuming the existence of band anticrossing point in the vicinity of the Fermi level, which acts as a magnetic monopole in the momentum space \cite{Fang_2003,Onoda_2006, Nagaosa_2010}.
In the present case, obtained $\sigma_{xy}^A$ in the WFM phase is less sensitive to pressure and consistent with
$e^2/(ha)=$ 440 $\Omega^{-1}$ cm$^{-1}$ using $a=8.841$ $\mathrm{\AA}$ \cite{Takemura_1988} at 3.2 GPa
[depicted with broken line in Fig. \ref{fig_5}(b)].
$\sigma_{xy}^A$ in the WFM phase of $\alpha$-Mn is comparable to those of Mn$_3$Sn ($\sim 150 \Omega^{-1}$ cm$^{-1}$), Mn$_3$Ge ($\sim 400 \Omega^{-1}$ cm$^{-1}$), and approximately half of that in elemental Fe ($\sim 1000 \Omega^{-1}$ cm$^{-1}$).
We also note that Fig. \ref{fig_5}(b) is quantitatively in agreement with the unified diagram of anomalous Hall physics \cite{Onoda_2006, Miyasato_2007, Nagaosa_2010, Liu_2018}.
At 3.8 GPa in Fig. \ref{fig_5}(b),
$\sigma_{xy}^A$ slightly deviates from the constant, which may relate on the cross-over from intrinsic to skew scattering mechanism.

\section{Conclusion}
In conclusion, a large anomalous Hall effect accompanied by the pressure-induced magnetic phase transition in $\alpha$-Mn was discovered,
which is the direct experimental evidence of Berry-curvature-associated anomalous Hall effect.
Despite the small spontaneous magnetization of $\sim$ 0.02 $\mu_B$/Mn, the anomalous Hall conductivity reaches 400-600 $\Omega^{-1}$ cm$^{-1}$ in the weak ferromagnetic phase under pressure, which is comparable to
non-collinear antiferromagnets Mn$_3$Sn and Mn$_3$Ge.
The anomalous Hall resistivity can be inverted by a miniscule switching magnetic field less than 10 mT,
and its magnitude can be widely controlled by external parameters.
The anomalous Hall conductivity is nearly independent of the relaxation time of impurity scattering,
which supports the dominant contribution of the Berry curvature in the weak ferromagnetic phase.
$\alpha$-Mn provides an elemental and tunable platform to unravel the large intrinsic Hall response by Berry curvature.
The present situation appears to be quite similar to that in Mn$_3$Sn and Mn$_3$Ge, in which a large intrinsic AHE emerges under a small but finite $M$.
In recent studies related to Mn$_3$Sn, the existence of the Weyl point in the momentum space \cite{Yang_2017, Kuroda_2017}
and a concept of cluster multipole moment \cite{Suzuki_2017} are proposed to explain the giant Hall response.
The specific origin of the non-zero contribution from the Berry curvature in pressurized $\alpha$-Mn remains an open question for future studies.

\begin{acknowledgments}
We thank J. Otsuki, Y. Yanagi, M. -T. Suzuki, H. Kusunose, and H. Harima for fruitful discussions and comments, and H. Ota for the X-ray single crystal structural analyses. This research was supported by JSPS KAKENHI Grant Numbers 18K03517 and 18H04323 (J-Physics). X-ray single crystal structural analyses were performed at the Division of Instrumental Analysis, Okayama University. 
\end{acknowledgments}

\bibliography{reference}

\end{document}